\documentclass[aps,pra,twocolumn,superscriptaddress,showpacs,amsmath,amssymb]{revtex4}
\usepackage[usenames,dvipsnames]{color}
\usepackage{epsf}
\usepackage{bm}
\usepackage{dcolumn}
\usepackage{latexsym}
\usepackage{amsmath}
\usepackage{amsfonts}
\usepackage{amssymb}
\usepackage{graphicx}
\usepackage[active]{srcltx}
\newcommand{\be}{\begin{equation}}\newcommand{\ee}{\end{equation}}
\newcommand{\bea}{\begin{eqnarray}}\newcommand{\eea}{\end{eqnarray}}
\newcommand{\brr}{\begin{array}}\newcommand{\err}{\end{array}}
\newcommand{\bit}{\begin{itemize}}\newcommand{\eit}{\end{itemize}}
\newcommand{\ben}{\begin{enumerate}}\newcommand{\een}{\end{enumerate}}

\newcommand{\bbm}{\begin{bmatrix}}\newcommand{\ebm}{\end{bmatrix}}
\newcommand{\ba}{\begin{array}}
\newcommand{\ea}{\end{array}}

\newtheorem{mydef}{Definition}
\newtheorem{Lemma}{Lemma}
\newtheorem{theorem}{Theorem}
\newcommand{\bd}{\begin{mydef}} \newcommand{\ed}{\end{mydef}}
\newcommand{\bthe}{\begin{theorem}} \newcommand{\ethe}{\end{theorem}}
\newcommand{\ble}{\begin{Lemma}} \newcommand{\ele}{\end{Lemma}}

\newcommand{\dr}{\mathrm{d}}

\definecolor{darkred}{rgb}{.8,0,0}

\definecolor{darkblue}{rgb}{0,0,.7}

\def\lan{\langle}
\def\lf{\left}

\def\non{\nonumber}\def\pa{\partial}\def\ran{\rangle}

\def\ri{\right}
\def\al{\alpha}\def\bt{\beta}\def\ga{\gamma}\def\Ga{\Gamma}
\def\de{\delta}

\def\om{\omega}

\def\1{{_{1}}}\def\2{{_{2}}}

\newcommand{\ide}{1\hspace{-1mm}{\rm I}}

\def\noHe0{:\;\!\!\;\!\!:H_e(0):\;\!\!\;\!\!:}
\def\noHm0{:\;\!\!\;\!\!:H_\mu(0):\;\!\!\;\!\!:}

\def\lan{\langle}
\def\lf{\left}

\def\non{\nonumber}
\def\pa{\partial}\def\ran{\rangle}

\def\ri{\right}

\def\al{\alpha}\def\bt{\beta}\def\ga{\gamma}
\def\Ga{\Gamma}\def\de{\delta}

\def\om{\omega}

\def\1{{_{1}}}\def\2{{_{2}}}

\def\A{{_{A}}}

\begin{document}

\title{Generalized generating functional for mixed-representation Green's functions: \\A quantum mechanical approach}

\author{Massimo~Blasone}
\email{blasone@sa.infn.it}
\affiliation{Dipartimento di Fisica, Universit\`a di Salerno, Via Giovanni Paolo II, 132 84084 Fisciano, Italy \& INFN Sezione di Napoli, Gruppo collegato di Salerno, Italy}
\author{Petr Jizba}
\email{p.jizba@fjfi.cvut.cz}
\affiliation{FNSPE, Czech Technical
University in Prague, B\v{r}ehov\'{a} 7, 115 19 Praha 1, Czech Republic\\}
\affiliation{ITP, Freie Universit\"{a}t Berlin, Arnimallee 14,
D-14195 Berlin, Germany}
\author{Luca~Smaldone}
\email{lsmaldone@sa.infn.it}
\affiliation{Dipartimento di Fisica, Universit\`a di Salerno, Via Giovanni Paolo II, 132 84084 Fisciano, Italy \& INFN Sezione di Napoli, Gruppo collegato di Salerno, Italy}
%
\vspace{3mm}

\begin{abstract}
When one tries to take into account the non-trivial vacuum structure of Quantum Field Theory, the standard functional-integral tools such as generating
functionals or transitional amplitudes, are often quite inadequate for such purposes. Here we propose a generalized generating
functional for Green's functions which allows to easily distinguish among a continuous set of vacua that are mutually connected via unitary canonical
transformations. In order to keep our discussion as simple as possible, we limit ourselves to Quantum Mechanics where the generating functional
of Green's functions is constructed by means of phase-space path integrals. The quantum-mechanical setting allows to accentuate the main logical steps involved without embarking on  technical complications such as renormalization or inequivalent representations that should otherwise be addressed in the full-fledged Quantum Field Theory.
We illustrate the inner workings of the  generating functional obtained by discussing
Green's functions among vacua that are mutually connected via translations and dilatations.
Salient issues, including connection with Quantum Field Theory, vacuum-to-vacuum transition amplitudes  and perturbation expansion in the vacuum parameter are also briefly discussed.
\end{abstract}

\maketitle

\section{Introduction}
%

Canonical transformations play a fundamental r\^{o}le in classical mechanics~\cite{Goldstein}, however,
their r\^{o}le in quantum physics is typically less significant~\cite{And,Umezawa,BJV}.
This status quo can be in part ascribed to the Groenewold--van~Hove ``no-go'' theorem~\cite{Groe,VH1951} which states that there exists a one-to-one correspondence between classical symplectic transformations and unitary transformations of quantum theory {\em only} when the generating function is at most quadratic, i.e., in the case of linear canonical transformations.
On the other hand, the linear canonical transformations disguised in the form of the Bogoliubov--Valatin transformations~\cite{Bog,Perelomov,Umezawa,BJV}  are central both in quantum mechanics (QM) and in quantum field theory (QFT). From the modern particle-physics and condensed-matter theory point of view, it is desirable to formulate the issues related to canonical transformations in the language of path integrals (PIs) as those often  provide  the  easiest  route  to  the  derivation  of perturbative expansions and serve as an excellent framework for
(both numerical and analytical) nonperturbative analysis~\cite{Schu1982,Kle,ItzZub,Atland,ZJ}



%

Our particular focus here will be on systems where the vacuum state is not invariant under canonical transformations.
This issue may be studied in its own right (e.g., in connection with semiclassical QM or theory of generalized coherent states)
but our primary motivation is dictated by prospective applications in QFT.
There, the   problem of the non-invariance of the vacuum state is particularly pressing because ensuing vacuum states typically belong to different (unitarily inequivalent) Hilbert spaces~\cite{BJV,Umezawa,Miransky,ItzZub}.
This situation shows up, e.g., in quantum systems with spontaneous symmetry breaking (SSB)~\cite{sewell,Miransky,BJSII}, in cases where renormalization issues are relevant~\cite{BJV,UTK,Haag,BJSII} or in
the study of flavor mixing both in flat~\cite{BJV,Mixing} and curved backgrounds~\cite{Blasone:2015bea}. The latter point has lead recently to phenomenologically relevant correction to the standard neutrino oscillation formula~\cite{BHV99}.
Another pertinent context where the multiple-vacuum structure plays an important r\^{o}le, is in study of time-dependent backgrounds in the ADS/CFT duality. There, the propagator among inequivalent vacua at different times predicts various non-trivial phenomenological effects such as cosmological particle creation~\cite{BalLevNaq}.


The PI (or better, its field-theoretic extension -- functional integral) treatment of QFT systems with a multiple vacuum structure was firstly studied 
in Ref.~\cite{MatPapUme}. There it was shown that the multiple vacua can be taken into account by using the so-called $\epsilon$-term prescription  (not to be mistaken with the Feynman--Stuckelberg $\epsilon$ prescription) for the generating functional of Green's functions. This allowed to fix a particular representation of the canonical commutation relations (CCR) and discuss, for instance, SSB in some systems. However, because of technical difficulties related to the analytic continuation of $\epsilon$, the method is seldom used in practice.
Further considerations have been pursued in the literature rather sparsely and in very specific contexts, see e.g., Refs.~\cite{AraIke,Torre,TelNog,BJSII}.

The aim of the present paper is to investigate the r\^{o}le of mixed-representation Green's functions.
To this end, we ask ourselves the following question:
 {\em Is it possible, in general, to construct a (generalized) generating functional for Green's functions which carries information about different vacua?}
%
Here the issue of the generalized generating functional (GGF) is addressed in the context of QM, where we show that the
aforestated question is answered affirmatively. We choose to work within QM, instead of QFT, for two basic reasons.
First, QM deals with systems with a finite number of degrees of freedom. In these cases, the Stone--von Neumann uniqueness theorem~\cite{Neu} ensures
that, for QM systems, the CCR, which govern the algebraic structure of
observables, admit only one irreducible Hilbert space representation. This is not true in QFT, where an infinite number of degrees of freedom  must be considered~\cite{Umezawa,BJV}. This simple looking fact has
far-reaching  consequences. The point is that in QFT there exists a separation between Heisenberg  fields, in terms of which the dynamical equations are written and asymptotic ({\em in}- or {\em out}-) fields which are directly related with the observed degrees of freedom~\cite{Umezawa,LSZ,StWi,BogLog}. At the same time, the Haag theorem~\cite{StWi,Haag} states that the Heisenberg  fields and {\em in}- (or {\em out}-) fields belong to unitarily inequivalent representations of CCR. This brings about technical complications related to the renormalization~\cite{UTK}. Since renormalization unnecessary obscures logical reasoning involved in the our construction of the  generalized generating functional, we prefer, for clarity's sake, to stick to QM setting.
Second, on the QM level there is a number of exact results for mixed-representation Green's functions that can be obtained via conventional canonical quantization.
These will serve as a gauge to which our conclusions (and hence internal consistency of the GGF obtained) can be compared.


The structure of the paper is as follows: In Section~\ref{sec1} we introduce the phase-space PIs and show how they imprint a choice of the Weyl--Heisenberg (WH) representation.
In order to keep track of various representations involved we derive a generalized generating functional for the mixed-representation Green's functions.
To put some flesh on the aforesaid generating functional, we discuss, in Section~\ref{pitrans}, an example of Green's functions with different vacua connected through spatial translations.
Despite its simplicity, this example represents proof-of-concept that the mixed-representation correlators can be systematically treated in practical situations including prospective QFT applications.
To bolster our exposition, we apply in Section~\ref{secdil} the generalized generating functional to a more challenging case, namely to the case of dilatations (scale and phase transformations). At the level of annihilation and creation operators, the latter correspond  to Bogoliubov--Valatin transformations~\cite{BJV,Miransky,Bog}.
Our discussion of dilatations traces a number of subtle issues related to the operator-ordering and provides resolution that coincides (for selected examples) with  canonical-quantization results.
For generic Hamiltonians we show how perturbative analysis in terms of the WH-vacuum parameter can be systematically carried out.
Finally, Section~\ref{SEc8} summarizes our results and discusses possible
extensions of the present work. For the reader's convenience, the paper is supplemented with two appendices which clarify some finer technical details needed in the main text.

\section{Canonical transformations and Green's functions in QM}\label{sec1}

Our starting point will be the phase-space PI representation of the evolution kernel $\lan q_f,t_f|q_i,t_i \ran$ which reads as
\be \label{kernzero}
\lan q_f, t_f|q_i, t_i \ran \ =\ \int^{q(t_f) = q_f}_{q(t_i) = q_i} \mathcal{D} q \,  \mathcal{D} p  \ e^{i S[p,q]} \, .
\ee
Here
\be
S[q,p]\ =\ \int^{t_f}_{t_i} \dr t \lf[p(t) \,  \dot{q}(t)\ -\ H(q(t),p(t))\ri] \, ,
\ee
is the phase-space action integral. For a future convenience we assume that  $\hbar =1$.
We also introduce the evolution kernel with source terms as
\be
\lan q_f,t_f|q_i,t_i \ran_{J_q,J_p} \ = \  \int^{q(t_f) = q_f}_{q(t_i) = q_i} \mathcal{D} q \,  \mathcal{D} p  \ e^{i S[p,q; \, J_q, J_p]} \, ,
\ee
where the new action integral has the form
\bea \non
S[p,q; \, J_q, J_p] &=& \int^{t_f}_{t_i} \dr t \lf[\ri. p(t) \dot{q}(t)\ -\ H(q(t),p(t)) \\
&-& J_q(t)q(t)\ -\ J_p(t)p(t)\lf.\ri] \, .
\eea
Here $J_q$ and $J_p$ are two auxiliary (Schwinger-type) currents with compact supports.

Algebraically, operators $\hat{q}(t)$ and $\hat{p}(t)$ form an irreducible representation of the Weyl--Heisenberg
algebra ${\mathcal{W}}_1$
\begin{eqnarray}
[\hat{q}(t), \hat{p}(t)] \ = \ i \hat{\ide}\, , \,\,\, [\hat{q}(t), \hat{\ide}] \ = \ [\hat{p}(t), \hat{\ide}] \ = \ 0\, .
\end{eqnarray}
%
Corresponding generalization from one to $N$ degrees of freedom constitutes the WH algebra ${\mathcal{W}}_N$.
The problem of classifying the representations of ${\mathcal{W}}_N$ is addressed by the Stone--von Neumann theorem~\cite{BJV,Perelomov}
which states that all irreducible unitary representations of the WH algebra ${\mathcal{W}}_N$ for any finite $N$ are unitarily equivalent.
This means that all irreducible representations of CCRs constitute an equivalent description of a given QM system.
The passage among such representations is mediated by an unitary operator
\begin{eqnarray} \label{unitgen}
\hat{G}_\al(t) \ = \ \exp\lf[i \al {K}\lf(\hat{q}(t),\hat{p}(t)\ri)\ri] , \qquad \alpha \in \mathbb{R}\, ,
\end{eqnarray}
where ${K}$   is by Stone's theorem some self-adjoint operator constructed from $\hat{q}$ and $\hat{p}$ operators.
With this, the coordinates and canonically conjugate momenta are transformed through an unitary mapping
\bea \label{transform1}
\hat{q}(t;\al) & \equiv & \hat{G}^{\dagger}_\al(t)\,  \hat{q}(t) \, \hat{G}_\al(t) \, , \\ [2mm]
\hat{p}(t;\al) & \equiv & \hat{G}^{\dagger}_\al(t)\,  \hat{p}(t) \, \hat{G}_\al(t) \, , \label{transform2}
\eea
where representation with $\alpha = 0$ is considered as a referential or fiducial representation. In spite of a mathematical equivalence, the
freedom in the choice of $\alpha$ can be conveniently used (and often is) to simplify various computational steps in QM --- most notably in the eigenvalue problem.
On the other hand, the choice of the representation
in QFT is much more than just a convenient mathematical trick.  In fact, since the Stone--von Neumann theorem
does not apply in QFT, different choices of a representation correspond to different physical realizations of the quantum system
and further elaboration is needed~\cite{BJV,Umezawa,ItzZub,UTK}.

To proceed let us first observe that, in general, $\hat{G}_\al$  does not leave the vacuum $|0\ran$ of the fiducial representation invariant. To see this, we define the vacuum state in
the new representation as
$ |0(\al,t)\ran \ \equiv \ \hat{G}_\al^{\dagger}(t) |0\ran$.
The overlap between the two vacua (vacuum persistence amplitude) is described by the transition amplitude
\bea
 \label{vacvac} \lan 0|0(\al,t)\ran\ =\ \lan 0|\hat{G}_\al^{\dagger}(t)|0\ran\, .
\eea
The latter plays a particularly important r\^{o}le in QFT, because there it provides a simple handle on the unitary (in)equivalence of the representations considered~\cite{BenHat}.
In order to compute (\ref{vacvac}) within the PI framework it is perhaps the simplest to start with a kernel in a generic representation labeled by  the index $\al \in \mathbb{R}$, i.e.
\bea
&&\mbox{\hspace{-12mm}}\lan q_f(\al),t_f|q_i(\al),t_i \ran  	\non \\
&& = \ \int^{q(t_f;\al) = q_f(\al)}_{q(t_i;\al) = q_i(\al)} \mathcal{D} q(\al) \,  \mathcal{D} p(\al)  \ e^{i S[p,q;\al]} \, ,
\label{II.10a}
\eea
where
\be
S[q,p; \al]\ =\ \int^{t_f}_{t_i} \dr t \lf[p(t;\al) \dot{q}(t;\al) - H(q(t;\al),p(t;\al))\ri] \, .
\ee
With the help of (\ref{II.10a}) one can also write down  the Feynman--Matthews--Salam formula~\cite{Schu1982,Kle,Landsman}
\begin{widetext}
\be \label{timeord2}
 \lan q_f(\al),t_f|T[q(t_n;\al)\ldots q(t_1;\al)]|q_i(\al),t_i \ran \ =\ \int^{q(t_f;\al) = q_f(\al)}_{q(t_i;\al) =
 q_i(\al)} \mathcal{D} q(\al)\,  \mathcal{D} p(\al)\, q(t_n;\al)\ldots q(t_1;\al)  \ e^{i S[p,q;\al]} \, ,
\ee
\end{widetext}
($T[\ldots]$ denotes the time-ordering symbol). Eq.~(\ref{timeord2}) holds  because the usual proof of the Feynman--Matthews--Salam formula does not invoke the choice of the representation at any stage.

We now introduce the $q$--ordering as
\begin{equation}
{\cal{O}}^q\left[ e^{i{K}(\hat{p}(\al;t),\hat{q}(\al;t))} \right] \ \equiv \
\sum_{k,l=0}^{\infty} K_{kl} \ \hat{q}^k(\al;t)\hat{p}^l(\al;t)\, .
\label{qorder}
\end{equation}
The latter orders the operator in such a way that all $\hat{q}(\al;t)$'s are on the left and $\hat{p}(\al,t)$'s are on the right. 
It should be stressed that the q-ordered operator $\mathcal{O}^q$ is equal to the original operator $\mathcal{O}$,  but it is written in such a way that $\hat{q}$'s and $\hat{p}$'s appear q-ordered. 
Note that any extra correction due to non-commutativity of $\hat{p}(\al;t),$ and $\hat{q}(\al;t)$ are   included in the coefficients $K_{kl}$.
Along the same lines we can define the ``classical'' $q$--ordering as
\begin{eqnarray} \non
&&\mbox{\hspace{-12mm}} {\mathcal{O}}^q_{cl}\!\left[e^{i {K}(p(\al;t),q(\al;t))} \right] \\
&&\ =\frac{\langle q(\al),t|{\mathcal{O}}^q\left[e^{i {K}(\hat{p}(\al;t),\hat{q}(\al;t))}\right]|p(\al),t \rangle}{\langle q(\al),t|p(\al),t \rangle} \, ,
\label{qcorder}
\end{eqnarray}
which proves to be important in the following considerations.
Having different vacua, we can now  define different sets of correlation functions. In particular, one can consider
\bea \label{green1} && \hspace{-0.8cm}i \mathcal{G}_{0}(t'-t) \ \equiv \ i \mathcal{G}_{0 0}(t'-t)\ =\ \lan 0|T\lf[\hat{q}(t') \hat{q}(t) \ri]| 0 \ran \, ,
\\[2mm]\label{green2}
&& \hspace{-0.8cm}i\mathcal{G}_{0 \al}(t'-t)\ =\ \lan 0|T\lf[\hat{q}(t';\al) \hat{q}(t;\al) \ri]| 0
\ran \, ,
\\[2mm]\label{green3}
&& \hspace{-0.8cm}i \mathcal{G}_{\bt 0}(t'-t)\ =\ \lan 0(\bt,t)|T\lf[\hat{q}(t') \hat{q}(t) \ri]| 0(\bt,t) \ran
\, ,
\\[2mm]\label{green4}
&& \hspace{-0.8cm}i\mathcal{G}_{\bt \bt}(t'-t) \ =\ \lan 0(\bt,t)|T\lf[\hat{q}(t';\bt) \hat{q}(t;\bt) \ri]| 0(\bt,t) \ran \, .
\eea\\

Although~\eqref{green1} can easily be recognized as a (causal) Green's function, in Appendix~A we prove that also the correlation functions~(\ref{green2})-(\ref{green4}) are Green's functions.

Correlators~\eqref{green1} and \eqref{green2} can be evaluated in a standard way. By starting from the matrix element~\eqref{timeord2},
we may write
\begin{widetext}
\be \label{trickT}
\lan q_f(\al),t_f|T[\hat{q}(t_n;\al)\ldots\hat{q}(t_1;\al)]|q_i(\al),t_i \ran \ =\ \sum_{n,m}\lan q_f(\al),t_f|n\ran \lan n|T[\hat{q}(t_n;\al)\ldots\hat{q}(t_1;\al)]| m \ran \lan m|q_i(\al),t_i \ran \, ,
\ee
\end{widetext}
where on the right-hand-side we have inserted the resolution of unity in terms of the energy eigenstates $|n\ran$.
We now want to take the limits $t_i \rightarrow -\infty$ and $t_f \rightarrow \infty$. Due to the oscillatory nature of the time evolution operator $e^{-i\hat{H}t}$ the limit does not make sense. In this connection we should, however, recall that by the spectral theorem,  $e^{-i\hat{H}z}$ with $z \in \mathbb{C}$ is strongly continuous for $\mbox{Im} z \leq 0$ and, in the present case, we are interested in the continuous boundary value  of an analytic function. To stay on the safe ground, we let $t_i \rightarrow -\infty$ and $t_f \rightarrow \infty$ in the complex plane, along a line with a small negative slope rather than real axis~\footnote{Rotation of the complex time axis back to the real axis will be done after outlined limits are performed.}.  Then, in the outlined limits, only the ground state contribute to (\ref{trickT}), which becomes
%
%
\bea
&& \hspace{-1.4cm}\lan q_f(\al),t_f|T[\hat{q}(t_n;\al)\ldots\hat{q}(t_1;\al)]|q_i(\al),t_i \ran \non \\[2mm]
&&=  \ \lim_{\substack{t_f \rightarrow +\infty \\ t_i \rightarrow -\infty}} e^{i E_0 (t_i-t_f)} \lan q_f(\al)|0\ran \non \\[2mm]
&& \times \  \lan 0| T[\hat{q}(t_n;\al)\ldots\hat{q}(t_1;\al)]|0\ran \lan 0 |q_i(\al)\ran \, .
\eea
This allows to identify the $n$-point correlation function with
\bea \label{firstgenstep}
&&\hspace{-0.6cm} \lan 0| T[\hat{q}(t_n;\al)\ldots\hat{q}(t_1;\al)]|0\ran \non \\[2mm]
&&\hspace{-0.4cm}= \lim_{\substack{t_f \rightarrow +\infty \\ t_i \rightarrow -\infty}} \frac{\lan q_f(\al),t_f|T[\hat{q}(t_n;\al) \ldots \hat{q}(t_1;\al)]|q_i(\al),t_i \ran }{\lan q_f(\al)|q_i(\al)\ran} \, ,\non \\
\eea
where we have used the identities
\bea
\lim_{t \rightarrow -\infty} |q(\al),t\ran &=& e^{i E_0 t}|0\ran \lan 0 |q(\al) \ran  \, , \label{3.20.aa} \\
\lim_{t \rightarrow +\infty} \lan q(\al),t| &=& e^{-i E_0 t} \lan q(\al) |0\ran \lan 0 | \, \, .
\label{id_1a}
\eea
By applying Eq.~\eqref{timeord2} we arrive at
\begin{widetext}
\be \label{tradformvac}
\lan 0| T[\hat{q}(t_n;\al)\ldots \hat{q}(t_1;\al)]|0\ran \ = \ \lim_{\substack{t_f \rightarrow +\infty \\ t_i \rightarrow -\infty}} \frac{\int^{q(t_f;\al)=q_f(\al)}_{q(t_i;\al)=q_i(\al)} \mathcal{D} q(\al)\,  \mathcal{D} p(\al) \, q(t_n;\al)\ldots q(t_1;\al)  \ e^{i S[p,q;\al]}}{\int^{q(t_f;\al)=q_f(\al)}_{q(t_i;\al)=q_i(\al)} \mathcal{D} q(\al)\,  \mathcal{D} p(\al) \ e^{i S[p,q;\al]}} \, .
\ee
\end{widetext}
In (\ref{id_1a}) we made an implicit choice of the fiducial vacuum state so that
\be
\hat{H}|0\ran \ = \ E_0 |0\ran \, .
\ee
On the other hand, it is evident that $|0(\al,t)\ran$ is not, in general, an eigenstate of $\hat{H}$. Indeed, from
\be
\hat{H}|0(\al,t)\ran \ = \ \hat{H} \hat{G}^{\dagger}_\al (t)|0\ran \, .
\ee
we see that $|0(\al,t)\ran$ is an eigenstate only if
\be
\lf[\hat{H}, \hat{G}_{\al}\ri]=0 \, ,
\ee
i.e., only when the transformation $\hat{G}_{\al}(t)$ is a symmetry of the problem. In this case, by using Eqs.~\eqref{transform1}-\eqref{vacvac}, it is easy to verify that Eqs.~\eqref{green1} and~\eqref{green2} coincide with Eqs.~\eqref{green4} and~\eqref{green3}, respectively, so our problem is completely solved. This is the reason of the success of the approach presented in Ref.~\cite{MatPapUme} in dealing with SSB.

Eq.~\eqref{tradformvac} can be recast into more compact form when the generating functional
\begin{widetext}
\be
\mathcal{Z}_{0  \al}[J_q]\ = \ \lim_{\substack{t_f \rightarrow +\infty \\ t_i \rightarrow -\infty}} \frac{\int^{q(t_f;\al)=q_f(\al)}_{q(t_i;\al)=q_i(\al)} \mathcal{D} q(\al)\,  \mathcal{D} p(\al) \ e^{i S[p,q;\al]+i \int^{+\infty}_{-\infty} J_q(t;\al) q(t;\al)}}{\int^{q(t_f;\al)=q_f(\al)}_{q(t_i;\al)=q_i(\al)} \mathcal{D} q(\al)\,  \mathcal{D} p(\al) \ e^{i S[p,q;\al]}}\, ,
\label{clagen}
\ee
\end{widetext}
is employed. Here the label $\al$ in  $J(t;\al)$ reminds that the current is coupled to $q(t;\al)$. With this we can equivalently rewrite (\ref{clagen}) in a succinct form as
\bea \label{hopeform1}
&&\mbox{\hspace{-10mm}}\lan 0| T(q(t_n;\al) \ldots q(t_1;\al))|0 \ran \non \\[2mm]
&&\mbox{\hspace{-8mm}} = \ \lf\{(-i)^n\frac{\de^n}{\de J_q(t_n;\al) \ldots \de J_q(t_1;\al)}\mathcal{Z}_{0 \, \al}[J_q]\ri\}_{J_q=0}\!\!.
\eea
When we set $n=2$ and $\al=0$ we get~\eqref{green1}, for $\al \neq 0$ we obtain~\eqref{green2}.

\begin{figure}[t]
\begin{center}
\includegraphics[width=7cm]{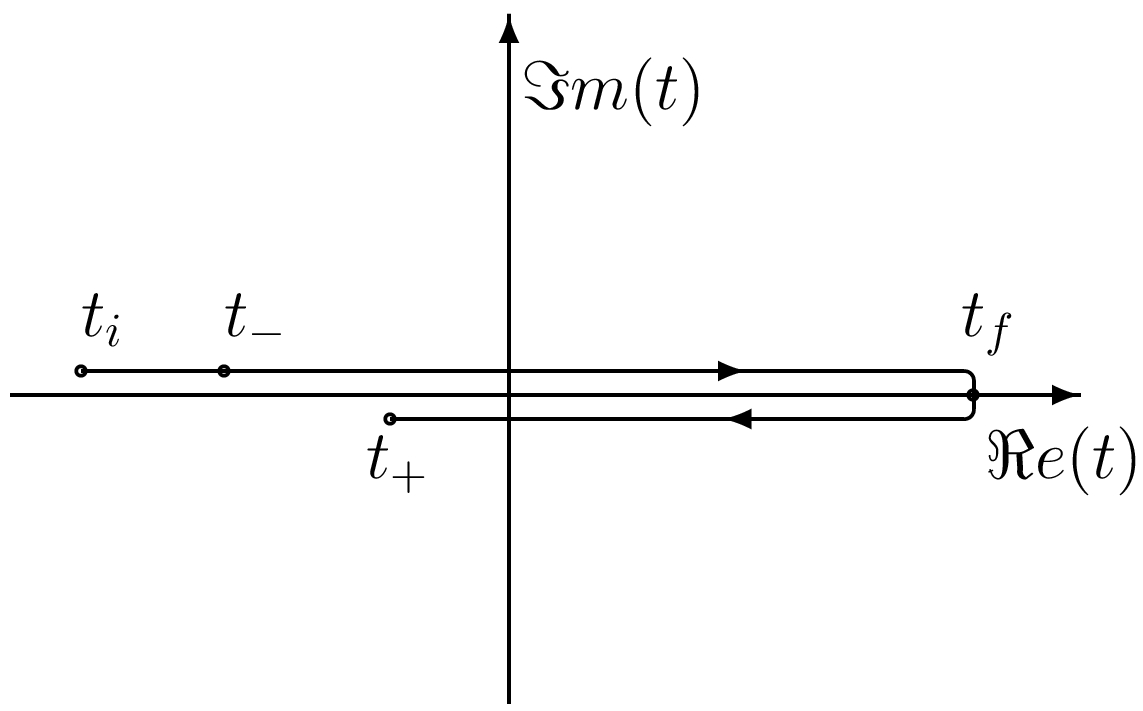}
\caption{The complex time contour $\mathcal{C}$ in the  Schwinger CTP formalism. Dots on the forward and the backward
branches of the contour denote discrete time points.}
\label{fig:closedtp}
\end{center}
\end{figure}

To compute Green's functions~\eqref{green3} and \eqref{green4} for general time-dependent $\hat{G}_{\al}(t)$, we must employ a different strategy than before.
In this case, we start with the expression
\bea
&&\mbox{\hspace{-11mm}} \lan 0(\bt,t_+)|T[\hat{q}(t_n;\al)\ldots\hat{q}(t_1;\al)]|0(\bt,t_-) \ran \non \\[2mm]
&&\mbox{\hspace{-5mm}} = \ \lan 0|\hat{G}_\bt(t_+)T[\hat{q}(t_n;\al)\ldots \hat{q}(t_1;\al)]\hat{G}_\bt^{\dagger}(t_-)|0 \ran \, ,
\label{mixvac0}
\eea
and assume that $t_+$ is bigger and  $t_-$ smaller than all time arguments involved in $T[\ldots]$. Following the same passages that brought us to Eq.~\eqref{firstgenstep}, we obtain
\begin{widetext}
\be  \label{mixvac}
\lan 0(\bt,t_+)|T[\hat{q}(t_n;\al)\ldots \hat{q}(t_1;\al)]|0(\bt,t_-) \ran\ =\  \lim_{\substack{t_f \rightarrow +\infty \\ t_i \rightarrow -\infty} }\frac{\lan q_f(\al),t_f|T[\hat{G}_\bt(t_+)\hat{q}(t_n;\al)\ldots\hat{q}(t_1;\al) \hat{G}_\bt^{\dagger}(t_-)] |q_i(\al),t_i \ran}{\lan q_f(\al),t_f|q_i(\al),t_i\ran} \, .
\ee
\end{widetext}
However, when deriving Eq.~\eqref{mixvac} all the operators $\hat{p}$ and $\hat{q}$ entering $\hat{G}_\bt$ have to be phrased in the $\alpha$-representation. Moreover, we suppose that the generator is $q$-ordered [see Eq.~\eqref{qorder}].
In this case we can readily rewrite (\ref{mixvac}) in the PI representation thanks to a simple generalization of Eq.~\eqref{timeord2}, namely
\begin{widetext}
\bea
&&\mbox{\hspace{-15mm}} \lan 0(\bt,t_+)|T[\hat{q}(t_n;\al)\ldots \hat{q}(t_1;\al)]|0(\bt,t_-) \ran  \ = \ \lan 0|T[ \hat{G}_\bt(t_+)\hat{q}(t_n;\al)\ldots \hat{q}(t_1;\al)\hat{G}_\bt^{\dagger}(t_-)]|0 \ran \non \\[2mm]
&& = \ \lim_{\substack{t_f \rightarrow +\infty \\ t_i \rightarrow -\infty} }\frac{\int^{q(t_f;\al)=q_f(\al)}_{q(t_i;\al)=q_i(\al)} \mathcal{D} q(\al)\,  \mathcal{D} p(\al) \, {\mathcal{O}}^q_{cl}\left[G_\bt(t_+)\ri] \, {\mathcal{O}}^q_{cl}\left[G_{-\bt}(t_-)\ri] \, q(t_n;\al) \ldots q(t_1;\al) \, \ e^{i S(p,q;\al)}}{\int^{q(t_f;\al)=q_f(\al)}_{q(t_i;\al)=q_i(\al)} \mathcal{D} q(\al)\,  \mathcal{D} p(\al) \ e^{i S(p,q;\al)}} \, .
\eea\\[2mm]
Note, in particular, the appearance of the $q$-ordered form ${\mathcal{O}}^q_{cl}[\ldots]$ of the generators $\hat{G}_\bt(t_+)$ and $\hat{G}_{-\bt}(t_-)$ which is a direct consequence of formulas (\ref{3.20.aa})-(\ref{id_1a}).  We have also used the simple fact that $\hat{G}^\dagger_{\bt}(t)  =  \hat{G}_{-\bt}(t)$.

We are thus naturally led to the following \emph{generalized generating functional}:
\bea \label{gentad1}
 \mathcal{Z}^{+ \, -}_{\bt \, \al}[J_q]   =  \left.\exp\left[ i f(\bt) K\lf(\frac{\de}{\de J_p(t_+;\al)},\frac{\de}{\de J_q(t_+;\al)}\ri)\right] \exp\left[i f(-\bt) K\lf(\frac{\de}{\de J_p(t_-;\al)},\frac{\de}{\de J_q(t_-;\al)}\ri)\right]\mathcal{Z}_{0\,\al}[J_q,J_p]\ri|_{J_p=0}\!\! ,
\eea
where
\be \label{gentad}
\mathcal{Z}_{0\,\al}[J_q,J_p]\ = \ \lim_{\substack{t_f \rightarrow +\infty \\ t_i \rightarrow -\infty}}\frac{\int^{q(t_f;\al)=q_f(\al)}_{q(t_i;\al)=q_i(\al)} \mathcal{D} q(\al)\,  \mathcal{D} p(\al) \ e^{i S(p,q;\al)+i \int^{t_f}_{t_i} \dr t
\lf[J_q(t) q(t;\al)+J_q(t) p(t;\al)\ri]}}{\int^{q(t_f;\al)=q_f(\al)}_{q(t_i;\al)=q_i(\al)} \mathcal{D} q(\al)\,  \mathcal{D} p(\al) \ e^{i S(p,q;\al)}} \, .
\ee
\end{widetext}
Instead of having simply $\beta$, the ordering issue forces us, in general, to consider a $c$-number function $f(\beta)$:
\begin{eqnarray}
{\mathcal{O}}^q_{cl}\left[\hat{G}_{\bt}(t)\right]  =  {\mathcal{O}}^q_{cl}\!\left[e^{i \beta K(\hat{p}(t),\hat{q}(t))} \right]  =  e^{i f(\beta)\, K\lf(q(t), p(t)\ri)}\, .
\end{eqnarray} 
This form results from the linear nature of considered canonical transformations (recall the Groenewold--van Hove no-go theorem). Some explicit examples of $f(\beta)$ will be derived in Sections~\ref{pitrans} and \ref{secdil}. With this in mind, we can write
\bea \label{hopeform}
&& \mbox{\hspace{-10mm}} \lan 0(\bt, t_+)| T(q(t_n; \al) \ldots q(t_1; \al))|0(\bt, t_-) \ran \nonumber \\[1mm]
&&  \mbox{\hspace{-5mm}}= \lf\{(-i)^n\frac{\de^n}{\de J_q(t_n;\al) \ldots \de J_q(t_1;\al)}\mathcal{Z}^{+  -}_{\bt\al}[J_q]\ri\}_{J_q=0} \! \! \!.
\eea
Note that, in order to obtain Eqs.~\eqref{green3}-\eqref{green4}, we have to set $t_+=t_-=t$ at the end. This, however, does not comply with the time ordering as defined in Eq.~\eqref{mixvac}. Of course, the implicit assumption in Eq.~\eqref{mixvac} is that all times involved belong to $\mathbb{R}$. By extending the time arguments to $\mathbb{C}$ so that $t_+ \mapsto t_+ -i\varepsilon$ ($\varepsilon$ is set to zero at the very end) we are naturally led to the Schwinger closed-time-path (CTP) formalism~\cite{SchCTP,Landsman} with the complex time-integration contour $\mathcal{C}$ shown in Fig.~\ref{fig:closedtp}. In this respect, the limit  $t_+=t_-=t$  should be understood so that calculations are done with non-zero $\varepsilon$ and only at the very end the sequence of limits $\lim_{t+ \rightarrow t_-=t} \lim_{\varepsilon \rightarrow 0}$ should be taken. Within this framework we can write
\begin{widetext}
\bea
&&\mbox{\hspace{-15mm}} \lan 0(\bt,t)|T[\hat{q}(t_n;\al) \dots \hat{q}(t_1;\al)]|0(\bt,t) \ran \ = \ \left. \lan 0|T_{\mathcal{C}}[ \hat{G}_\bt(t_+)\hat{q}(t_n;\al)\ldots \hat{q}(t_1;\al)\hat{G}_\bt^{\dagger}(t_-)]|0 \ran\right|_{t_+ \rightarrow t_-=t} \non \\[2mm]
&& =\ \lim_{\substack{t_f \rightarrow +\infty \\ t_i \rightarrow -\infty} }\frac{\left.  \int^{q(t_f;\al)=q_f(\al)}_{\mathcal{C} \ q(t_i;\al)=q_i(\al)} \mathcal{D} q(\al)\,  \mathcal{D} p(\al){\mathcal{O}}^q_{cl}\left[G_\bt(t_+)\ri]{\mathcal{O}}^q_{cl}\left[G_{-\bt}(t_-)\ri] \, q(t_n;\al)\ldots q(t_1;\al)  \ e^{i S(p,q;\al)}\right|_{t_+ \rightarrow t_- = t}}{\int^{q(t_f;\al)=q_f(\al)}_{q(t_i;\al)=q_i(\al)} \mathcal{D} q(\al)\,  \mathcal{D} p(\al) \ e^{i S(p,q;\al)}} \, .
\label{2.37.aa}
\eea
\end{widetext}

Green's functions~\eqref{green3},\eqref{green4} can be thus obtained as
\bea \label{green3bis}
&&\mbox{\hspace{-10mm}}i\mathcal{G}_{\bt  0}(t'-t) \non \\
&&\mbox{\hspace{-10mm}}= \lim_{t_+ \rightarrow t_- = t} \lf\{(-i)^2\frac{\de^2}{\de J_q(t') \de J_q(t)}\mathcal{Z}^{+ -}_{\bt  0}[J_q]\ri\}_{J_q=0} \!\! \!,
\\ [2mm]\label{green4bis}
&&\mbox{\hspace{-10mm}} i\mathcal{G}_{\bt  \bt}(t'-t)  \non \\
&&\mbox{\hspace{-10mm}}= \lim_{t_+ \rightarrow  t_- = t} \lf\{(-i)^2\frac{\de^2}{\de J_q(t';\bt) \de J_q(t;\bt)}\mathcal{Z}^{+  -}_{\bt  \bt}[J_q]\ri\}_{J_q=0}\!\!\!
.
\eea
We should emphasize that the limit cannot be, in general, exchanged with the functional derivatives, otherwise we could obtain erroneous results due to the fact that we implicitly work within the CTP formalism. For instance, Eqs.~\eqref{green3} and \eqref{green4} would be equal to Eqs.~\eqref{green2} and \eqref{green1}, respectively.
On the other hand, as we have already pointed out, this can happen only when $\hat{G}_{\al}(t)$ is a symmetry of the problem.

Note that, if $\beta$ is small and $f(\beta)=o(\beta)$, we can write a perturbative expansion in $\beta$ of Eq.~\eqref{gentad1}, where the leading-order in $\beta$ reads
\bea \label{firstord}
&&\mbox{\hspace{-9mm}}\mathcal{Z}^{+  -}_{\bt \al}[J_q]\ \approx\ \mathcal{Z}_{0 \al}[J_q]\non \\[2mm]
&&\mbox{\hspace{-9mm}}+\ i \bt \lf[ K\lf(\frac{\de}{\de J_p(t_+;\al)},\frac{\de}{\de J_q(t_+;\al)}\ri)\ri.\non \\[2mm]
&&\mbox{\hspace{-9mm}}-\ \lf.\lf.K\lf(\frac{\de}{\de J_p(t_-;\al)},\frac{\de}{\de J_q(t_-;\al)}\ri)\ri]\mathcal{Z}_{0  \al}
[J_q,J_p]\ri|_{J_p=0}\!\!.
\eea
Here the $0$-order contribution $\mathcal{Z}_{0 \al}[J_q]$ denotes the generating functional~\eqref{clagen}.

We now generalize our previous reasoning a bit and introduce yet another generating functional
\bea
&&\mbox{\hspace{-10mm}} \mathcal{Z}^{ +  -}_{\ga  \bt  \al}[J_q]\ = \  e^{i f(\ga) K\lf(\frac{\de}{\de J_p(t_+;\al)},\frac{\de}{\de J_q(t_+;\al)}\ri)} \non \\[2mm]
&&\mbox{\hspace{-5mm}}  \times \ \lf. e^{i f(-\bt) K\lf(\frac{\de}{\de J_p(t_-;\al)},\frac{\de}{\de J_q(t_-;\al)}\ri)}\mathcal{Z}_{0  \al}[J_q,J_p]\ri|_{J_p=0} \! .
\eea
From this, we find that the vacuum-to-vacuum transition amplitude~\eqref{vacvac} is simply
\bea \label{green3tris}
\lan 0| 0(\bt,t) \ran \ = \ \lim_{t_- \rightarrow t}\mathcal{Z}^{+  -}_{0  \bt  0}[0] \, .
\eea
Let us stress, once more, that in the QFT setting this transition amplitude would typically be zero due to non-existence  of the operator $K$ (the domain  of $K$ tends to zero in the long-wave limit). See, for instance,  Refs.~\cite{ItzZub,Umezawa,BJV,BJSII,Miransky} for detailed discussions of this issue.

\section{Translations} \label{pitrans}

We will now see how the GGF actually work, by calculating the  1- and 2-point correlation functions in mixed-representation for two simple cases.
In our first example we consider translations defined through the prescription
\be \label{deftrans}
\hat{q}(t;\al) \ =\ \hat{q}(t) \ + \ \al \, , \qquad \hat{p}(t;\al) \ = \ \hat{p}(t) \, .
\ee
The generator of this transformation is
\be
\hat{G}(\al;t)\ =\ \exp[-i \al \hat{p}(t)] \label{gentransl}\, .
\ee
We now consider the  Hamiltonian
\begin{eqnarray} \label{hamlin}
&&\mbox{\hspace{-12mm}}\hat{H}(\hat{q}(t;\al),\hat{p}(t;\al))\nonumber \\[2mm]
&& =\ \frac{\hat{p}^2(t;\al)}{2}\ +\ \frac{\hat{q}^2(t;\al)}{2}\ +\ {P}\lf(\hat{q}(t;\al)\ri) \, ,
\end{eqnarray}
where the potential has the form
\be \label{defp}
{P}\lf(\hat{q}(t;\al)\ri) \ = \ - \ \al \,\hat{q}(t;\al)\ +\  \frac{\al^2}{2} \, .
\ee
In the QFT context the Hamiltonian (\ref{hamlin})  would correspond to the so-called van-Hove model~\cite{vanHove}, describing an infinite 
chain of linear harmonic oscillators (LHOs) subject to an external force.

Transformation~\eqref{deftrans} brings Eq.~\eqref{hamlin} in the LHO form
\be \label{harmosc}
\hat{H}(\hat{q}(t),\hat{p}(t))\ =\ \frac{\hat{p}^2(t)}{2} \ + \ \frac{\hat{q}^2(t)}{2} \, ,
\ee
where
$m= \om = 1$.

Our starting point here is the evolution kernel
\bea
&&\mbox{\hspace{-5mm}} \lan q_f(\al),t_f| q_i(\al),t_i\ran_{J_q \, J_p} \non \\[2mm]
&&\mbox{\hspace{-3mm}} = \ \int^{q(t_f;\al) = q_f(\al)}_{q(t_i;\al)  = q_i(\al)} \mathcal{D} q(\al)\,  \mathcal{D} p(\al)\,  e^{i \int_{t_i}^{t_f} \dr t \, p(t;\al)\dot{q}(t;\al)} \non \\[2mm]
&&\mbox{\hspace{-3mm}} \times \ e^{-i \int_{t_i}^{t_f} \dr t\lf[H(q(t;\al),p(t;\al))\ -\ J_p (t;\al) p(t;\al)\ -\ J_q (t;\al) q(t;\al)\ri]} \, . \nonumber \\
\label{alphakernel}
\eea
By performing the change of variables
\bea
q(t;\al) \ = \ q(t)\ +\ \al \, ,  \qquad
p(t;\al) \ = \ p(t) \, ,
\eea
Eq.~\eqref{alphakernel} reduces to
\bea \label{djker}
&&\mbox{\hspace{-5mm}} \lan q_f(\al),t_f| q_i(\al),t_i\ran_{J_q \, J_p} \non \\[2mm]
&&\mbox{\hspace{-3mm}}=\ \int^{q(t_f)=q_f \equiv q_f(\al)-\al}_{q(t_i;\al)=q_i \equiv q_i(\al)-\al} \mathcal{D} q\,  \mathcal{D} p \, e^{i \int_{t_i}^{t_f} \dr t \lf[p(t)\dot{q}(t)-H(q(t),p(t))\ri]} \non \\[2mm]
&&\mbox{\hspace{-3mm}} \times \ e^{i \int_{t_i}^{t_f} \dr t \ \lf[ J_p (t;\al) p(t)+J_q (t;\al) q(t)\ri]}e^{i \int_{t_i}^{t_f} \dr t J_q (t;\al) \al\ } \, .
\eea
Apart from the last exponential factor, the other pieces are now standard and we can now easily write down the GGF of Green's functions~\eqref{clagen} in the explicit form (cf. e.g., Ref.~\cite{BJV})
\bea
&&\mbox{\hspace{-7mm}}\mathcal{Z}_{0  \al}[J_q,J_p]\ = \ e^{-\frac{i}{2}\int^{+\infty}_{-\infty} \dr \tau \int^{+\infty}_{-\infty} \dr \tau'  J_q(\tau;\al)\mathcal{G}(\tau-\tau') J_q(\tau';\al)}\non \\[2mm]
&& \times\ e^{-i \int^{+\infty}_{-\infty} \dr \tau \int^{+\infty}_{-\infty} \dr \tau' J_q(\tau;\al)\pa_{\tau'}\mathcal{G}(\tau-\tau') J_p(\tau';\al)} \non \\[2mm]
&& \times\ e^{-\frac{i}{2} \int^{+\infty}_{-\infty} \dr \tau \int^{+\infty}_{-\infty} \dr \tau' J_p(\tau;\al)\mathcal{G}(\tau-\tau') J_p(\tau';\al)}\non \\[2mm]
&& \times \ e^{i \al \int^{+\infty}_{-\infty} \dr \tau J_q(\tau;\al)}\, .
\eea
Here $\mathcal{G}(t-t')$ represents the ordinary LHO Green's function, i.e.
\bea \label{greendef1}
\lf(-\pa^2_{t}-1\ri)\mathcal{G}(t-t')& = & \de(t-t')\, , \\[2mm]
\lf(-\pa^2_{t'}-1\ri)\mathcal{G}(t-t')& = & \de(t-t')\, . \label{greendef2}
\eea
Eqs.~\eqref{greendef1} and \eqref{greendef2} leave $\mathcal{G}(t-t')$ undefined up to an homogeneous solution.
A typical choice is to take $\mathcal{G}(t-t') \ =\ \mathcal{G}_0(t-t') $, where $\mathcal{G}_0(t-t')$ is defined by Eq.~\eqref{green1}
which coincides with the Feynman--St\"{u}ckelberg causal propagator
\begin{eqnarray} \label{hogf}
&&\mbox{\hspace{-10mm}}\mathcal{G}_0(t'-t) \ = \ \int \ \frac{\dr k}{2 \pi} \,  \frac{e^{-i k (t-t')}}{k^2-1+i \varepsilon}\non \\[2mm]
&&\mbox{\hspace{-5mm}} = \ -\frac{i}{2} \lf[\theta(t'-t) e^{-i(t'-t)} + \theta(t-t') e^{i(t'-t)}\ri]  .
\end{eqnarray}

Because we have no ordering problem here (the generator~\eqref{gentransl}, involves only $\hat{p}$), the function $f(\al)$, introduced in Eq.~\eqref{gentad} reduces to $f(\bt)  =  i \bt$. Hence the GGF~\eqref{gentad1} reduces to
\bea
&&\mbox{\hspace{-10mm}} \mathcal{Z}^{+  -}_{\bt \al}[J_q]  =   e^{-\bt \frac{\de}{\de J_p(t_+;\al)}} \lf. e^{\bt \frac{\de}{\de J_p(t_-;\al)}}\mathcal{Z}_{0  \al}[J_q,J_p]\ri|_{J_p=0} \!\! \!.
\eea
By employing the simple identity
\be
\exp\left(\rho \pa_x\right)f(x)\ =\ f(x+ \rho) \, ,
\ee
where $\pa_x \equiv \frac{\partial}{\partial x}$, we find that
\begin{widetext}
\bea
&&\mbox{\hspace{-12mm}}\mathcal{Z}^{+  -}_{\bt  \al}[J_q] \ = \ e^{-\frac{i}{2}\int^{+\infty}_{-\infty} \dr \tau \int^{+\infty}_{-\infty} \dr \tau'   J_q(\tau;\al)\mathcal{G}_0(\tau-\tau')J_q(\tau';\al)\ +\ i \al \int^{+\infty}_{-\infty} \dr \tau J_q(\tau;\al)}\non \\[2mm]
&&\mbox{\hspace{-5mm}} \times  \ \lf[ e^{-\bt \frac{\de}{\de J_p(t_+;\al)}}e^{-i \int^{+\infty}_{-\infty} \dr \tau \int^{+\infty}_{-\infty} \dr \tau' J_q(\tau;\al)\pa_{\tau'}\mathcal{G}_0(\tau-\tau') J_p(\tau';\al)} \ri. \non \\[2mm]
&&\mbox{\hspace{-5mm}} \times\  \lf.  e^{-\frac{i}{2} \int^{+\infty}_{-\infty} \dr \tau \int^{+\infty}_{-\infty} \dr \tau' J_p(\tau,\al)\mathcal{G}_0(\tau-\tau') J_p(\tau';\al)}e^{-\frac{\bt^2}{4}-i \bt\int^{+\infty}_{- \infty} \dr \tau \, \mathcal{G}_0(\tau-t_-) J_p(\tau;\al)-i \bt \int^{+\infty}_{-\infty} \dr \tau \,  J_q(\tau;\al)\pa_{t_-}\mathcal{G}_0(\tau-t_-)} \ri]_{J_p=0} .
\eea
\end{widetext}
%
%
%
If we take the functional derivative and set $J_p=0$, the GGF may be written in the form
\bea
&&\mbox{\hspace{-10mm}} \mathcal{Z}^{+  -}_{\bt  \al}[J_q] \ = \ e^{-\frac{i}{2}\int^{+\infty}_{-\infty} \dr \tau \int^{+\infty}_{-\infty} \dr \tau' J_q(\tau;\al)\mathcal{G}_0(\tau-\tau')J_q(\tau';\al)}\non \\[2mm] \label{gengf}
&& \times \ e^{i \int^{+\infty}_{-\infty} \dr \tau J_q(\tau;\al)\lf[\al-\bt\pa_{t_-}\mathcal{G}_0(\tau-t_-)+\bt\pa_{t_+}\mathcal{G}_0(\tau-t_+)\ri]} \non \\[2mm]
&& \times\ e^{-\frac{\bt^2}{4}-i \bt^2 \mathcal{G}_0(t_+-t_-)}\, .
\eea
In the same way we find that
\bea
&&\mbox{\hspace{-10mm}}  \mathcal{Z}^{+  -}_{0  \bt  0}[J_q]\ =\ e^{-\frac{i}{2}\int^{+\infty}_{-\infty} \dr \tau \int^{+\infty}_{-\infty} \dr \tau'   J_q(\tau)\mathcal{G}_0(\tau-\tau')J_q(\tau')}\non \\[2mm]
&&\mbox{\hspace{-6mm}}  \times \ e^{i \bt \int^{+\infty}_{-\infty} \dr \tau J_q(\tau)-\frac{\bt^2}{4}- i \bt \int^{+\infty}_{-\infty} \dr \tau J_q(\tau)\pa_{t_-}\mathcal{G}_0(\tau-t_-) }\, ,
\eea
so, we can  evaluate the vacuum-vacuum transition amplitude~\eqref{vacvac} to be
\be
\lan 0|0(\bt, t)\ran\ =\ \lim_{t_- \rightarrow t} \mathcal{Z}^{+  -}_{0 \bt  0}[0]\ =\ \exp\lf[-\frac{\bt^2}{4}\ri] \, .
\ee
In principle,  we can now derive any $n$-point correlation function from the GGF~\eqref{gengf}. However, for simplicity's sake we limit ourselves to 1-point and 2-point correlation functions. In particular, our aim is to compute Green's functions~\eqref{green2}-\eqref{green4}. By taking a functional derivative of the GGF~\eqref{gengf} with respect to $J_q(t_1,\al)$, we find
\bea
&&\mbox{\hspace{-5mm}} i \frac{\de\mathcal{Z}^{+  -}_{\bt \al}[J_q]}{\de J_q(t_1,\al)}\nonumber \\[2mm]
&&\mbox{\hspace{-5mm}} =\ \lf[ \int^{+\infty}_{-\infty} \dr \tau  J_q(\tau;\al)\mathcal{G}_0(\tau-t_1)  -  \al + \bt\pa_{t_-}\mathcal{G}_0(t_1-t_-)\ri.  \non \\[2mm]
&&\mbox{\hspace{-5mm}} -   \lf.  \bt\pa_{t_+}\mathcal{G}_0(t_1-t_+)\ri] \mathcal{Z}^{+  -}_{\bt  \al}[J_q]\, .
\eea
Setting $J_q = 0$  and $t_1=t$ we get
\bea
&&\mbox{\hspace{-12mm}} \lan 0(\bt,t_+)|\hat{q}(t;\al)|0(\bt;t_-)\ran \ = \ \lf[\al-\bt\pa_{t_-}\mathcal{G}_0(t-t_-)\right. \nonumber \\[2mm]
&&\mbox{\hspace{19mm}}  +\ \left. \bt\pa_{t_+}\mathcal{G}_0(t-t_+)\ri]\mathcal{Z}^{+ -}_{\bt \al}[0]\,  ,
\eea
where
\be
\mathcal{Z}^{+  -}_{\bt \al}[0] \ = \ \exp\left[-\frac{\bt^2}{4}-i \bt^2 \mathcal{G}_0(t_+-t_-)\right] \, .
\ee
In the case when $\bt=\al=0 $ we recover the expected result $\lan 0|\hat{q}(t)|0\ran=0$. When $\bt=0$ but $\al \neq 0$ we obtain
\be \label{exvalbeta}
\lan 0|\hat{q}(t;\al)|0\ran\ =\ \al \, .
\ee
If $\al=0$ but $\bt \neq 0$ we find
\bea
&&\mbox{\hspace{-9mm}} \lan 0(\bt,t_+)|\hat{q}(t)|0(\bt;t_-)\ran \non \\[2mm]
&&\mbox{\hspace{-7mm}} =\ \lf[\bt\pa_{t_+}\mathcal{G}_0(t-t_+)\ \ -\bt\pa_{t_-}\mathcal{G}_0(t-t_-)\ri]\mathcal{Z}^{+  -}_{\bt  \al}[0] \, .
\label{65a}
\eea
To explicitly evaluate expression (\ref{65a}), we employ the fact [see Eq.~\eqref{hogf}] that
\bea \label{greenpartic1}
\mathcal{G}_0(t-t_-)& = & -\frac{i}{2} e^{-i(t-t_-)} \, , \\[2mm] \label{greenpartic2}
\mathcal{G}_0(t-t_+)& =& -\frac{i}{2} e^{i(t-t_+)} \, .
\eea
Thanks to Eqs.~\eqref{greenpartic1} and \eqref{greenpartic2}, we can write
\bea
&&\mbox{\hspace{-14mm}} \lan 0(\bt,t_+)|\hat{q}(t)|0(\bt;t_-)\ran \non \\[2mm]
&&\mbox{\hspace{-4mm}}= \ -\frac{ \bt}{2} \lf(e^{-i(t-t_-)}\ +\ e^{i(t-t_+)}  \ri) \mathcal{Z}^{+  -}_{\bt  \al}[0] \, .
\eea
At this stage we can safely set $t_+ = t_- = t'$ and observe that
\be \label{onepoint1}
\lan 0(\bt,t')|\hat{q}(t)|0(\bt;t')\ran\ =\ - \bt \cos(t-t')\, .
\ee
In the same way we find
\be \label{onepoint}
\lan 0(\bt,t')|\hat{q}(t;\bt)|0(\bt;t')\ran\ =\ \bt \lf[1\ -\ \cos(t-t')\ri]\, .
\ee
%

Let us now consider the 2-point correlation functions. These can be written as
\begin{widetext}
\bea
&&\mbox{\hspace{-5mm}} \lan 0(\bt,t_+)|T\lf[\hat{q}(t_2;\al)\hat{q}(t_1;\al)\ri]|0(\bt,t_-)\ran  \ = \ (-i)^2 \lf.\frac{\de^2\mathcal{Z}^{+  -}_{\bt  \al}[J_q]}{\de J_q(t_2;\al)\de J_q(t_1;\al)}\ri|_{J_q=0} \non \\
&&\mbox{\hspace{-5mm}} = \ \lf\{i\mathcal{G}_0(t_2-t_1)\ +\ \lf[\ri.\al-\bt\pa_{t_-}\mathcal{G}_0(t_1-t_-)\ + \ \bt\pa_{t_+}\mathcal{G}_0(t_1-t_+)\lf.\ri]\lf[\al-\bt\pa_{t_-}\mathcal{G}_0(t_2-t_-)\ +\ \bt\pa_{t_+}\mathcal{G}_0(t_2-t_+)\ri] \ri\} \mathcal{Z}^{+  -}_{\bt \al}[0]\, . \non \\[1mm]
\label{3.69.a}
\eea
\end{widetext}
If we set $\bt = \al = 0$ we recover the usual Feynman--St\"{u}ckelberg causal propagator~\eqref{green1}. On the other hand, if we set  $\bt = 0, \al \neq 0$ and put $t_2=t' \, , \, t_1=t$, we obtain the Green's function~\eqref{green2} in the explicit form
%
\be \label{simpgreen}
i \mathcal{G}_{0 \, \al}(t'-t) \ =\  i\mathcal{G}_0(t'-t)\ +\ \al^2 \, .
\ee
If $\al = 0$ and $\bt \neq 0$ we get
\begin{widetext}
\bea
\lan 0(\bt,t_+)|T\lf[\hat{q}(t_2)\hat{q}(t_1)\ri]|0(\bt,t_-)\ran & =& \lf\{ i\mathcal{G}_0(t_2-t_1)+\ \lf[\bt\pa_{t_+}\mathcal{G}_0(t_1-t_+)\ - \ \bt\pa_{t_-}\mathcal{G}_0(t_1-t_-)\ri] \ri. \non\\[2mm]
 && \times \ \lf. \lf[\bt\pa_{t_+}\mathcal{G}_0(t_2-t_+)\ -\ \bt\pa_{t_-}\mathcal{G}_0(t_2-t_-)\ri] \ri\} \mathcal{Z}^{+  -}_{\bt \al}[0]  \, .
\eea
\end{widetext}
With the help of Eqs.~\eqref{greenpartic1},\eqref{greenpartic2} this can be rewritten as
\bea
&&\mbox{\hspace{-7mm}} \lan 0(\bt,t_+)|T\lf[\hat{q}(t_2)\hat{q}(t_1)\ri]|0(\bt,t_-)\ran \non \\[2mm]
&& = \ \lf\{ i\mathcal{G}_0(t_2-t_1)+\frac{\bt^2}{4}\lf[e^{-i(t_1-t_-)}\ +\ e^{i(t_1-t_+)}\ri]\ri. \non \\[2mm]
&&\times \ \lf. \lf[e^{-i(t_2-t_-)}\ +\ e^{i(t_2-t_+)}\ri] \ri\} \mathcal{Z}^{+  -}_{\bt  \al}[0]  \, .
\eea
Setting $t_2 = t'$ and $t_-=t_+=t_1=t$ we get  Green's function~\eqref{green3} in the explicit form 
\be \label{regreen3}
i\mathcal{G}_{\bt \, 0}(t'-t) \ =\  i\mathcal{G}_0(t'-t)\ +\ \bt^2 \cos (t-t') \, .
\ee
It is easy to see that the difference   $ i\mathcal{G}_{\bt \, 0}(t'-t) - i\mathcal{G}_0(t'-t)$ represents  an homogeneous solution of Eq.~\eqref{evolgreen1}. This implies that both Green's functions differ only by the choice of boundary conditions (cf. Appendix~A). 

If in Eq.~(\ref{3.69.a}) we set $\bt = \al$ we get the Green's function~\eqref{green4}
\be \label{regreen4}
i \mathcal{G}_{\bt \, \bt}(t'-t)\ =\ i\mathcal{G}_0(t'-t) \, .
\ee
Note, that the difference $ i\mathcal{G}_{\bt \, \bt}(t'-t) - i\mathcal{G}_{0 \, \bt}(t'-t)$  is a solution of Eqs.~\eqref{newgreen1},\eqref{newgreen2} for $t \neq t'$, i.e., an homogenous solution, as again shown in Appendix A.

\section{Dilatations} \label{secdil}

The second example which we discuss here is that of dilatations:
\bea \label{dil1}
&&\hat{q}(t;\al)\  = \ \hat{G}_{\al}^{-1}(t)\, \hat{q}(t)\, \hat{G}_{\al}(t) \ =\
e^{\al} \, \hat{q}(t)\, , \\[2mm]
&&\hat{p}(t;\al)\ =\ \hat{G}_{\al}^{-1}(t)\, \hat{p}(t)\, \hat{G}_{\al}(t) \ =\
e^{-\al} \, \hat{p}(t)\, .
\label{dil2}
\eea
In this case the generator $\hat{G}_{\al}(t)$ has the form
\be \label{gendil}
\hat{G}_{\al}(t) \ = \ \exp\left\{-i \al \lf[\hat{q}(t) \, \hat{p}(t)-\frac{i}{2}\ri]\right\} \, .
\ee
We will consider the Hamiltonian
\be \label{dilham}
\hat{H}(\hat{q}(t;\al),\hat{p}(t;\al))\ =\ \frac{\hat{p}^2(t;\al)^2 e^{2 \al}}{2} \ +\ \frac{\hat{q}^2(t;\al)^2 e^{-2 \al}}{2} \, .
\ee
As in Section~\ref{pitrans}, this reduces to the LHO  form~\eqref{harmosc} after the transformation~\eqref{dil1}-\eqref{dil2} is employed.
This fact would be important in the QFT perturbation approach, where the LHO-like part of the action defines the free asymptotic field and ensuing perturbation Green's function.

Proceeding as in Section~\ref{pitrans} one can derive the generating functional $\mathcal{Z}_{0  \al}[J_q,J_p]$:
\bea \non
&&\mbox{\hspace{-6mm}} \mathcal{Z}_{0 \, \al}[J_q,J_p] \ = \ e^{-\frac{i}{2} \int^{+\infty}_{-\infty} \dr \tau \dr \tau' e^{2 \al }J_q(\tau;\al)\mathcal{G}_0(\tau-\tau') J_q(\tau';\al)}\\[2mm]
&& \times \ e^{-i \int^{+\infty}_{-\infty} \dr \tau \dr \tau' J_q(\tau;\al)\pa_{\tau'}\mathcal{G}_0(\tau-\tau') J_p(\tau';\al)} \non \\[2mm]
&&  \times  \ e^{-\frac{i}{2} \int^{+\infty}_{-\infty} \dr \tau  \dr \tau' e^{-2 \al}J_p(\tau;\al)\mathcal{G}_0(\tau-\tau') J_p(\tau';\al)}\, . \label{partdil}
\eea
The corresponding GGF~\eqref{gentad1} reads now as
\begin{widetext}
\be \label{genpartdil}
\mathcal{Z}^{+  -}_{\bt  \al}[J_q] \ =\ \left. \exp\left\{i f(\bt) \frac{\de}{\de J_p(\tau_+;\al)}\frac{\de}{\de J_q(\tau_+;\al)}\right\} \exp\left\{i f(-\bt) \frac{\de}{\de J_p(\tau_-;\al)}\frac{\de}{\de J_q(\tau_-;\al)}\right\}\mathcal{Z}_{0 \, \al}[J_q,J_p]\ri|_{J_p=0} \, .
\ee
\end{widetext}
First of all, we notice that the $c$-number factor present in Eq.~\eqref{gendil} disappears in the definition~\eqref{genpartdil}. It remains to determine the function $f(\bt)$. To this end we use the relation~\cite{celeghini1996quantum}
\be
\lf(\hat{q}(t)\hat{p}(t)\ri)^n\ =\ \sum^n_{k=1}(-i)^{n-k}S^{(k)}_{n}\hat{q}^k(t)\hat{p}^k(t) \, ,
\ee
where $S^{(k)}_{n}$ are Stirling numbers of the second kind, defined by the recurrence relation~\cite{AbrSteg, celeghini1996quantum}
\be
S^{(k)}_{n+1}\ =\ k S^{(k)}_{n}\ +\ S^{(k-1)}_{n} \, .
\ee
Thanks to the identity
\be
\sum^\infty_{n=1}\sum^n_{k=1}\ = \ \sum^\infty_{k=1}\sum^\infty_{n=k} \, ,
\ee
and to the generating relation
\be
\frac{1}{k !} (e^\bt-1)^k \ = \ \sum^\infty_{n=k} S^{(k)}_{n} \frac{\bt^k}{k !} \, ,
\ee
we find that
\be
{\cal{O}}^q\left[e^{-i\bt \hat{q}(t) \, \hat{p}(t)} \right]\ =\  \, \sum^\infty_{k=0} i^k \hat{q}^k(t)\hat{p}^k(t) \frac{(e^{-\bt}-1)^k}{k!} \, .
\ee
With this we can also easily check that
\be \label{Oqcl}
{\cal{O}}^q_{cl} \left[e^{-i\bt \hat{q}(t) \, \hat{p}(t)} \right]\ = \  \exp\left[i(e^{-\bt}-1)\, q(t)p(t)\right] \, ,
\ee
and thus  identify  $f(\bt) = 1-e^{-\bt}$.

Before we embark on the computation of Eq.~\eqref{genpartdil}, we derive first the vacuum-to-vacuum transition amplitude~\eqref{vacvac} by using the formula~\eqref{strongformula1} from Appendix~B. In the present case we obtain
\be 	\label{gammacoeff}
\Ga\ =\ 1 \ + \ 4 f^2(-\bt) \left(a b \ -\ c^2 \right)\ -\ 4 i f(-\bt) c \, ,
\ee
where $\Ga$ is defined in Eq.~(\ref{gammacoeff1}).

The coefficients in $\Ga$ can be determined by inspecting Eq.~\eqref{partdil}. This gives
\bea
a &=& \frac{i}{2}\mathcal{G}_0(0)\, e^{-2 \bt} \ =\  \frac{1}{4}e^{-2 \bt} \, , \label{co1} \\[2mm]
b &=& \frac{i}{2}\mathcal{G}_0(0)e^{2 \bt}\ =\  \frac{1}{4} e^{2 \bt} \, , \label{co2} \\[2mm]
c &=& -\frac{i}{2}\pa_t\mathcal{G}_0(0)\ =\  -\frac{i}{4} \, . \label{co3}
\eea
The derivative term in Eq.~\eqref{co3} should be understood as
\bea
\pa_t\mathcal{G}_0(0) & = & \lim_{t'-t \rightarrow 0^+} \lan 0 |\hat{q}(t')\hat{p}(t)|0\ran \\
& = & \pa_t \lim_{t'-t \rightarrow 0^+} \lan 0 |\hat{q}(t')\hat{q}(t)|0\ran \, ,
\eea
because of the $q$-ordering prescription employed.
Thanks to Eq.~\eqref{co1} and \eqref{co3} we can write
\be\Ga\ =\ \frac{e^{2 \bt}+1}{2}\ =\ e^\bt  \cosh \bt \, .
\ee
Consequently, we obtain the vacuum-to-vacuum transition amplitude in the final form
\be
\lan 0 |0(\bt,t)\ran \ =\ \lim_{t_- \rightarrow t}\mathcal{Z}^{+  -}_{0  \bt  0}[0] \ =\  \frac{1}{\sqrt{\cosh \bt}} \, .
\ee
Let us now come back to Eq.~\eqref{genpartdil}. An exact evaluation would be rather long and not very illuminating. However, it is instructive to use the example of dilatations to illustrate how perturbation considerations work in practice. For small $\beta$ the GGF (\ref{firstord}) acquires the explicit form
\bea
&&\mbox{\hspace{-8mm}} \mathcal{Z}^{+  -}_{\bt  \al}[J_q]\ \approx \ \mathcal{Z}_{0  \al}[J_q]- i \bt\lf[ \frac{\de}{\de J_p(t_-;\al)}\frac{\de}{\de J_q(t_-;\al)} \ri. \non \\[2mm]
&& - \ \lf. \lf. \frac{\de}{\de J_p(t_+;\al)}\frac{\de}{\de J_q(t_+;\al)}\ri]\mathcal{Z}_{0  \al }[J_q,J_p]\right|_{J_p=0}\! ,
\eea
which reduces to
\begin{widetext}
\bea
&&\mbox{\hspace{-13mm}} \mathcal{Z}^{+  -}_{\bt \al}[J_q] \ \approx \ \exp\left[-\frac{i \,  e^{2 \al}}{2} \int^{+\infty}_{-\infty} \dr \tau \int^{+\infty}_{-\infty} \dr \tau' J_q(\tau;\al)\mathcal{G}_0(\tau-\tau') J_q(\tau';\al)\right]\non \\[2mm]
&&\mbox{\hspace{-10mm}}\times \  \lf[1-i \, \bt \, e^{2 \al} \int^{+\infty}_{-\infty} \dr \tau \int^{+\infty}_{-\infty} \dr \tau' J_q(\tau;\al)\lf[\pa_{t_+}\mathcal{G}_0(\tau-t_+)\mathcal{G}_0(\tau'-t_+)-\pa_{t_-}\mathcal{G}_0(\tau-t_-)\mathcal{G}_0(\tau'-t_-) \ri]J_q(\tau';\al)\ri]\!\! .
\label{98a}
\eea
To the leading order in $\beta$ this can be rewritten in the simple compact form as
\bea
\mathcal{Z}^{+  -}_{\bt \al}[J_q] \ = \ \exp\left[-\frac{i \,  e^{2 \al}}{2} \int^{+\infty}_{-\infty} \dr \tau \dr \tau' J_q(\tau;\al)\mathcal{F}^{+  -}_\bt(\tau-\tau') J_q(\tau';\al)\right] \, ,
\label{99aa}
\eea
where
\bea
\mbox{\hspace{-5mm}} \mathcal{F}^{+  -}_\bt(t-t') & = & \mathcal{G}_0(\tau-\tau')  \ + \ 2 \bt \lf[\ri. \pa_{t_+}\mathcal{G}_0(\tau-t_+)\mathcal{G}_0(\tau'-t_+) \ + \ \pa_{t_-}\mathcal{G}_0(\tau-t_-)\mathcal{G}_0(\tau'-t_-)\lf.\ri] \non \\[2mm]
&=& \ \mathcal{G}_0(\tau-\tau') +\ \frac{i \bt }{2}  \lf[e^{-i(t_+-\tau)}e^{-i(t_+-\tau')}\ri. \ +\ \lf. e^{i(t_- -\tau)}e^{i(t_--\tau')}\ri]
\, .
\label{100aa}
\eea
\end{widetext}
Thanks to GGF~(\ref{99aa}), we can now evaluate the leading-order correlation functions.
%
One can check that all one-point correlation functions are identically zero.
It is also clear that, to the leading-order, we have
\be
\lim_{t_+ \rightarrow t_-=t}\mathcal{F}^{+  -}_\bt(t-t') \ = \ \mathcal{G}_{\bt  0}(t-t') \, ,
\ee
and so, thanks to Eq.~\eqref{100aa}, we get
\bea \label{tiringres}
\hspace{-0.5cm}i \mathcal{G}_{\bt \, \al}(t'-t)  \ = \ e^{2 \al}  \lf[ i\mathcal{G}_0(t'-t)\ - \ \bt \cos(t'-t)\ri]  .
\eea
If $\bt  =  \al  =  0$ we recover the standard Green's function~\eqref{green1}. If $\bt=0$ we get the Green's function~\eqref{green2}:
\bea
i \mathcal{G}_{0 \, \al}(t'-t')  \ = \ e^{2 \al}  i\mathcal{G}_0(t'-t) \, .
\eea
If $\al=0$ we get the Green's function~\eqref{green3}
\bea
\hspace{-0.5cm}i \mathcal{G}_{\bt \, 0}(t'-t)  \ = \  i\mathcal{G}_0(t'-t)\ - \ \bt \cos(t'-t) \,   .
\eea
Finally, if $\al \ =\  \bt$ we get the Green's function~\eqref{green4}
\bea
\hspace{-0.5cm}i \mathcal{G}_{\bt \, \bt}(t'-t)  \ = \ e^{2 \bt}  \lf[ i\mathcal{G}_0(t'-t)\ - \ \bt \cos(t'-t)\ri]  .
\eea
Let us note, that in the leading-order approximation the aforementioned Green's functions differ from each other only in boundary terms (homogenous solutions).

\section{Conclusions} \label{SEc8}

In the present article we have derived a generalized generating functional for  mixed-representation Green's functions in the phase-space PI representation. Although, at a mathematical level, the origin of the  mixed-representation correlation functions is quite clear, being related to a certain class of canonical transformations, at the QM level they are not often considered. Exceptions are provided by semiclassical QM  semiclassical QM and theory of generalized (Perelomov-type) coherent states.
A context in which they turn out to be truly relevant is  QFT, where the existence of unitarily inequivalent Fock spaces~\cite{BJV,Umezawa} allows for different vacuum states. As a consequence, in QFT the choice of a vacuum state and ensuing correlation functions has  non-trivial phenomenological implications. Relevant examples of this are, e.g., flavor mixing~\cite{Mixing}, renormalization~\cite{BJV,Col,KadBay}, continuous phase transitions~\cite{sewell,ZJ,KadBay} and Thermo Field Dynamics~\cite{Umezawa}.

An important upshot of our analysis is the finding that  the respective members of such class  of correlation functions differ among themselves only by boundary terms (homogeneous solutions).
Indeed, the fact that the Green's functions studied in the present work, differ from the conventional ones in the specification of boundary conditions, implies the necessity at a physical level of careful definition of asymptotic states and ensuing perturbative expansion. This is indeed the case of mixing, in which two distinct sets of asymptotic states exist~\cite{Mixing}. Other important cases include QFT on curved backgrounds~\cite{Birrell} or unstable particles.

We have shown that such boundary conditions can be included in an elegant and unified way in the generalized generating functional of Green's functions. There the boundary terms are reflected in the parametrization of the interacting term epitomized by the generating function $K$. It is interesting to notice that during our analysis we were naturally led to the so-called Schwinger closed-time-path formulation of Green's functions which typically appears when discussing out-of-equilibrium interacting many-body systems. We have illustrated the power of the generalized generating functional obtained on two explicit examples, namely translations (van-Hove model) and dilatations.




An obvious direction for a future extension of the present work would be in realm of QFT functional integrals (FIs). There one requires a covariant formulation in terms of Lagrange functions instead of Hamiltonians.  One of the important advantages of the FI formalism is its manifestly Lorentz covariance which is lost when using Hamiltonian FIs (as treated here in the context of PIs) and closer connection with some of the most pressing issues of modern  high-energy particle physics.

Another interesting route to follow would be to find out the connection with the generalized coherent-state (\`{a} la Perelomov) PIs and the present phase-space-type of GGF.


\begin{acknowledgments}
P.J.  was  supported  by the Czech  Science  Foundation Grant No. 17-33812L. P.J. and L.S. were also in part supported by the U.S. Army RDECOM - Atlantic Grant no. W911NF-17-1-0108.
\end{acknowledgments}

\section*{Appendix A}
In this Appendix we present some general considerations about the Green's functions~\eqref{green2}-\eqref{green4} used in the main body of the text.

Let us first verify that  ${\cal G}_{\bt \, 0}(t,t')$  is a Green's
function, i.e. it satisfies
\bea \label{evolgreen1}
 \hspace{-0.5cm}L(\partial_{t})\, {\cal G}_{\bt \, 0}(t,t')\ =\
L(\partial_{t'})\, {\cal G}_{\bt \, 0}(t,t') \ =\
\de(t-t')\, .
\eea
Here $L(\partial_t)$ represents the differential operator such that $ L(\partial_{t})
\hat{q}(t)=0$ is the  equation of motion for
$\hat{q}(t)$ in Heisenberg's picture.
%
%
To do so, we need to show that ${\cal G}_{\bt \, 0}(t,t')={\cal G}_{\bt \, 0}(t-t')$.
This follows easily from the following chain of reasonings 
\begin{widetext}
\bea
{\cal G}_{\bt \, 0}(t,t') & = & \theta(t-t') \lan 0| e^{i \hat{H} t/\hbar}
\hat{G}_\bt(0) e^{-i \hat{H} t/\hbar}\, \hat{q}(t') \, \hat{q}(t) \, e^{i
\hat{H} t/\hbar} \, \hat{G}^{-1}_\bt(0) \, e^{-i \hat{H} t/\hbar} | 0 \ran
\non \\[2mm]
&&+ \ \theta(t'-t) \lan 0| e^{i \hat{H} t/\hbar} \, \hat{G}_\bt(0) \, e^{-i
\hat{H} t/\hbar}\,  \hat{q}(t) \, \hat{q}(t')\,  e^{i \hat{H}
t/\hbar} \, \hat{G}^{-1}_\bt(0) \, e^{-i \hat{H} t/\hbar} | 0 \ran
\non \\[2mm]
& = &  \theta(t-t') \lan 0| \hat{G}_\bt(0) \, \hat{q}(t'-t) \, \hat{q}(0) \,
\hat{G}^{-1}_\bt(0)| 0 \ran \ + \ \theta(t'-t) \lan 0| \hat{G}_\bt(0) \hat{q}(0)
\hat{q}(t'-t) \,
\hat{G}^{-1}_\bt(0) | 0 \ran
\non \\[2mm]
& = &  {\cal G}_{\bt \, 0}(t-t')\, .
\eea
\end{widetext}
Since the considered equations of motion are quadratic and linear in time, see Eqs.~\eqref{greendef1},\eqref{greendef2}, we have
\bea
&&\mbox{\hspace{-11mm}}L(\partial_{t})\, {\cal G}_{\bt\, 0}(t-t') \ = \ L(\partial_{t'})\, {\cal
G}_{\bt \, 0}(t-t')\,  ,
\eea
as it should be according to Eq.~(\ref{evolgreen1}). Note that the $\delta$-function on the right-hand-side of (\ref{evolgreen1}) comes from the derivative of $\theta(t-t')$ in the definition of the two-point function.
It should be stressed that ${\cal G}_{\bt \, 0}(t-t')$ differs from the usual propagator $\mathcal{G}_0(t-t')$, which obeys the same equation of motion, in the choice of boundary conditions.

However, it can be seen that $\mathcal{G}_{0 \, \al}(t-t')$ is not, in general, a Green's function in the same sense which we have seen above, but rather it is in the sense typically understood in QFT (see, e.g., Ref.~\cite{Col}), i.e. satisfying a infinite set of coupled equations among different-order correlation functions (so-called Schwinger--Dyson equations~\cite{ItzZub,Col,KadBay}). Let us consider, e.g., the simple example discussed in Section~\ref{pitrans}. Thanks to the Hamiltonian~\eqref{hamlin} we obtain the evolution equation
\be \label{evoltrans}
\pa^2_t\hat{q}(t; \al) \ + \ \hat{q}(t; \al) \ = \ \al \, .
\ee
Using Eq.~\eqref{evoltrans}, it can be easily verified that
\bea \label{newgreen1}
 &&\mbox{\hspace{-20mm}}(\pa^2_{t'} + 1 ) \mathcal{G}_{0 \, \al} \ + \ \lan 0 | T\lf[{P}'[\hat{q}(t';\al)] \hat{q}(t; \al) \ri]|0\ran\nonumber \\[2mm]  &&\mbox{\hspace{-15mm}}= \ \de(t'-t) \, , \\[2mm]
 &&\mbox{\hspace{-20mm}}(\pa^2_{t} + 1 ) \mathcal{G}_{0 \, \al} \ + \  \lan 0 | T\lf[\hat{q}(t';\al) {P}'[\hat{q}(t; \al)] \ri]|0\ran\nonumber \\[2mm]   &&\mbox{\hspace{-15mm}}= \ \de(t'-t) \, ,\label{newgreen2}
\eea
where ${P}[\hat{q}(t; \al)]$ was introduced in Eq.~\eqref{defp} and, formally
\be
{P}'[\hat{q}(\tau;\al)]=\frac{\de {P}[\hat{q}(\tau;\al)] }{\de \hat{q}(\tau;\al)}\, .
\ee
So, explicitly we have
\bea
&&\hspace{-0.7cm}(\pa^2_{t'} + 1 ) \mathcal{G}_{0 \, \al}\ -\ \al \lan 0 | \hat{q}(t; \al)|0\ran \ = \ \de(t'-t) \, ,\\
&&\hspace{-0.7cm}(\pa^2_{t} + 1 ) \mathcal{G}_{0 \, \al}\ -\ \al \lan 0 | \hat{q}(t';\al) |0\ran \ = \ \de(t'-t) \, ,
\eea
which relates 1-point functions with the propagator.

Following similar reasonings as above, one can show that $\mathcal{G}_{\bt \, \bt}(t-t')$ is a
Green's function in the above mentioned QFT sense. In the same manner one can prove that $\mathcal{G}_{\al\, \al}(t-t')$ should differ from $\mathcal{G}_{0 \, \al}(t-t')$ only in the choice of the boundary conditions. This point has important implications, for instance,  for setting up appropriate perturbation expansion in QFT. We will illustrating this issue more explicitly in our future work.

\section*{Appendix B}

Here we show an instructive toy model calculation of the GGF (\ref{genpartdil}). In doing so, 
we calculate factor $\Gamma$ and other quantities that are used in Section~IV.

Let us consider the expression:
\be
 e^{\rho \pa_x \pa_y}e^{-a x^2-b y^2+2cxy} \, ,
\ee
Now we perform the substitution
\bea
a x^2  \ - \ 2c x y  \ +\ b y^2  & = &  a \left( x  - \frac{c}{ a} y \right)^2 \ +\
y^2 \left(b-\frac{c^2}{a} \right) \non \\[2mm]
& = & a\xi^2 \ + \ h \gamma^2 \, ,
\eea
with $ \xi \ \equiv x \ -\ {c}{a}/ y\, , \;
\gamma
\equiv y\, , \; h \equiv b-{c^2}/{a} $. With this
\be
 \partial_x\partial_y \ =\
- \ \frac{c}{a}\partial_{\xi}^2\  + \ \partial_{\xi}\partial_{\gamma}\, ,
\ee
and hence
\be \label{loca}
 e^{\rho \pa_x \pa_y}e^{-a x^2-b y^2+2cxy}=e^{-\frac{\rho c}{a}\partial_{\xi}^2 + \rho \partial_{\xi}\partial_{\gamma}}e^{-a\xi^2 -h \gamma^2} \, .
\ee
We start evaluating
\be
e^{\bt \pa^2_\xi} \, e^{-a \xi^2}\ = \ \sum^{\infty}_{n=0}\frac{\bt^n}{n!}\pa^{2n}_\xi e^{-a \xi^2}  \, .
\ee
Using the Rodrigues formula for Hermite polynomials~\cite{GraRyz}, in the case of even degrees:
\be
\pa^{2n}_\xi e^{-a \xi^2}\ =\ a^{n} \, H_{2n} \, (\sqrt{a}\xi) \,  e^{-a \xi^2} \, ,
\ee
we get
\be
e^{\bt \pa^2_\xi} \, e^{-a \xi^2} \ =\  \sum^{\infty}_{n=0}\frac{\bt^n}{n!} \, a^{n} \, H_{2n}(\sqrt{a}\xi) \, e^{-a \xi^2}  \, .
\ee
It is useful to rewrite Hermite polynomials in the integral representation~\cite{GraRyz}:
\be \label{inthermrap}
H_m(x)\ =\ \frac{2^m}{\sqrt{\pi}} \int^{+\infty}_{-\infty} \dr t \ (x+i t)^m \, e^{-t^2} \, ,
\ee
so that
\bea
&& e^{\bt \pa^2_\xi} \, e^{-a \xi^2}\\ \non
&& =\  \frac{1}{\sqrt{\pi}}\int^{+\infty}_{-\infty} \dr t \sum^{\infty}_{n=0}\frac{\bt^n}{n!}a^{n}( \sqrt{a} \xi+it)^{2n} 2^{2n} e^{-t^2} e^{-a \xi^2}  \, .
\eea
Resumming the series, which now is a simple exponential series, we get
\be
e^{\bt \pa^2_\xi}\, e^{-a \xi^2}\ =\ \frac{1}{\sqrt{\pi}}\int^{+\infty}_{-\infty} \dr t  \ e^{-t^2+ 4 a \beta  \left(\sqrt{a} \xi+i t\right)^2} \,  e^{-a \xi^2} \, .
\ee
Performing the integral we thus have
\be \label{compres}
e^{\bt \pa^2_\xi} \, e^{-a \xi^2} \ =\ \frac{e^{-\frac{a \xi^2}{ 1+4 a \bt} }}{\sqrt{1+4 a \bt}} \, , \qquad \mbox{Re}\left(a\beta\right)\ > \ -\frac{1}{4} \, .
\ee
Putting $\bt\ =\ -(\rho c)/a$ we can rewrite Eq.\eqref{loca} as
\be
e^{\rho \pa_x \pa_y} \, e^{-a x^2-b y^2+2cxy} \ = \ \sqrt{\frac{g}{a}} \, e^{\rho \partial_{\xi}\partial_{\gamma}} \, e^{-h \gamma^2-g \xi^2} \, ,
\ee
where
\be
g\ =\ \frac{a}{1-4 \rho c} \, \, .
\ee
Expanding the first exponential, we have
\be
e^{\rho \partial_{\xi}\partial_{\gamma}} \, e^{-h \gamma^2-g \xi^2} \ =\ \sum^\infty_{n=0} \, \frac{\rho^n}{n!} \, \partial^n_{\xi}\partial^n_{\gamma} \, e^{-h \gamma^2-g \xi^2} \, .
\ee
Firstly, we concentrate on
\be
e^{\rho \partial_{\xi}\partial_{\gamma}} \, e^{-g \xi^2} \ =\ \sum^\infty_{n=0} \frac{\rho^n}{n!} \partial^n_{\xi} e^{-g \xi^2} \partial^n_{\gamma} \, .
\ee
Applying once more Rodrigues' formula
\be
\pa^{n}_\xi e^{-a \xi^2} \ =\ a^{\frac{n}{2}}(-1)^n H_{n}(\sqrt{a}\xi) \, e^{-a \xi^2} \, ,
\ee
we get
\be
e^{\rho \partial_{\xi}\partial_{\gamma}} \, e^{-g \xi^2}\ =\ e^{-g \xi^2} \sum^\infty_{n=0} \frac{(-\rho \sqrt{g})^n}{n!}  H_n(\sqrt{g}\xi) \partial^n_{\gamma} \, .
\ee
Using the integral representation~\eqref{inthermrap} we find
\be
e^{\rho \partial_{\xi}\partial_{\gamma}} \, e^{-g \xi^2} \ =\ e^{-g \xi^2} \, e^{-2 g \rho \xi \pa_\ga-\rho^2 g \pa^2_\ga} \, .
\ee
Our final results can be thus written as
\be \label{resder}
e^{\rho \partial_{\xi}\partial_{\gamma}} \, e^{-h \gamma^2-g \xi^2} = \frac{e^{-g \xi^2-\frac{h \ga^2-4 g \rho h \xi \ga+4 h g^2 \rho^2 \xi^2}{ 1-4 g \rho^2 h}}}{\sqrt{1\ -\ 4 \rho^2  \left(a b \ -\ c^2 \right)-4 \rho c}} \, .
\ee
In the original variables $x$ and $y$ this reads as
\begin{eqnarray} \label{strongformula}
&&\mbox{\hspace{-12mm}}\exp\left(\rho \pa_x \pa_y\right)\exp\left( -a x^2-b y^2+2cxy\right) \nonumber \\[2mm]
&&\mbox{\hspace{-12mm}} =  \frac{1}{\sqrt{\Ga}}\exp\left[\frac{-a x^2-b y^2+2cxy+4 xy \rho(ab-c^2)}{\Ga}\right]\!,
\end{eqnarray}
where
\be 	\label{gammacoeff1}
\Ga\ = \ 1\ -\ 4 \rho^2  \left(a b \ -\ c^2 \right)-4 \rho c \, .
\ee
In analogy with (\ref{genpartdil}) we should set $x = y = 0$. With this we consequently obtain
\be \label{strongformula1}
\lf.e^{\rho \pa_x \pa_y}e^{-a x^2-b y^2+2cxy}\ri|_{x,y=0} \ =\ \frac{1}{\sqrt{\Ga}} \, .
\ee
%

\end{document}